# Biological data integration using Semantic Web technologies

Pasquier C

Phone: +33 492 07 6947

Fax: +33 492 07 6432

Email: claude.pasquier@unice.fr

Institute of Signaling, Developmental Biology & Cancer

CNRS - UMR 6543, University of Nice Sophia-Antipolis

Parc Valrose, 06108 NICE cedex 2, France.

## Summary

Current research in biology heavily depends on the availability and efficient use of information. In order to build new knowledge, various sources of biological data must often be combined. Semantic Web technologies, which provide a common framework allowing data to be shared and reused between applications, can be applied to the management of disseminated biological data. However, due to some specificities of biological data, the application of these technologies to life science constitutes a real challenge.

Through a use case of biological data integration, we show in this paper that current Semantic Web technologies start to become mature and can be applied for the development of large applications. However, in order to get the best from these technologies, improvements are needed both at the level of tool performance and knowledge modeling.

## Keywords

# Introduction

Biology is now an information-intensive science and research in genomics, transcriptomics and proteomics heavily depend on the availability and the efficient use of information. When data were structured and organized as a collection of records in dedicated, self-sufficient databases, information was retrieved by performing queries on the database using a specialized query language; for example SQL (Structured Query Language) for relational databases or OQL (Object Query Language) for object databases. In modern biology, exploiting the different kinds of available information about a given topic is challenging because data are spread over the World Wide Web (Web), hosted in a large number of independent, heterogeneous and highly focused resources.

The Web is a system of interlinked documents distributed over the Internet. It allows access to a large number of valuable resources, mainly designed for human use and comprehension. Actually, hypertext links can be used to link anything to anything. By clicking a hyperlink on a Web page, one frequently obtains another document which is related to the clicked element (this can be a text, an image, a sound, a clip, etc). The relationship between the source and the target of a link can have a multitude of meanings: an explanation, a traduction, a localization, a sell or buy order, etc. Human readers are capable of deducing the role of the links and are able to use the Web to carry out complex tasks. However, a computer cannot accomplish the same tasks without human supervision because Web pages are designed to be read by people, not by machines.

Hands-off data handling requires moving from a Web of documents, only understandable by humans, to a Web of data in which information is expressed not only in natural language, but also in a format that can be read and used by software agents, thus permitting them to find, share and integrate information more easily [1]. In parallel with the Web of data, which is focused primarily on data interoperability, considerable international efforts are ongoing to develop programmatic interoperability on the Web with the aim of enabling a Web of programs [2]. Here, semantic descriptions are applied to processes, for example represented as Web Services [3]. The extension of both the static and the dynamic part of the current Web is called the Semantic Web.

The principal technologies of the Semantic Web fit into a set of layered specifications. The current components are the Resource Description Framework (RDF) Core Model, the RDF Schema language (RDF schema), the Web Ontology Language (OWL) and the SPARQL query language for RDF. In this paper, these languages are designed with the acronym SWL for Semantic Web Languages. A brief description of these languages, which is needed to better understand this paper, is given below.

The Resource Description Framework (RDF) model [2] is based upon the idea of making statements about resources. A RDF statement, also called a triple in RDF terminology is an association of the form *(subject, predicate, object)*. The subject of a RDF statement is a resource identified by a Uniform Resource Identifier (URI) [3]. The predicate is a resource as well, denoting a specific property of the subject. The object, which can be a resource or a string literal, represents the value of this property. For example, one way to state in RDF that "*the human gene BRCA1 is located on chromosome 17*" is a triple of specially formatted strings: a subject denoting "*the human gene BRCA1*", a predicate representing the relationship "*is located on*", and an object denoting "*chromosome 17*". A collection of triples can be represented by a labeled directed graph (called RDF graph) where each vertex represents either a subject or an object and each edge represents a predicate.

RDF applications sometimes need to describe other RDF statements using RDF, for instance, to record information about when statements were made, who made them, or other similar information (this is sometimes referred to as "provenance" information). RDF provides a built-in vocabulary intended for describing RDF statements. A description of a statement using this vocabulary is called a reification of the statement. For example, a reification of the statement about the location of the human gene BRCA1 would be given by assigning the statement a URI (such as http://example.org/triple12345) and then, using this new URI as the subject of other statements, like in the triples *(http://example.org/triple12345, specified_in, "human assembly_N")* and *(http://example.org/triple12345, information_coming_from, "Ensembl_database")*.

RDF Schema (RDFS) [4] and the Web Ontology Language (OWL) [5] are used to explicitly represent the meanings of the resources described on the Web and how they are related. These specifications, called ontologies, describe the semantics of classes and properties used in Web documents. An ontology suitable for the example above might define the concept of Gene (including its relationships with other concepts) and the meaning of the predicate "*is located on*". As stated by John Dupré in 1993 [4], there is no unique ontology. There are multiple ontologies which each models a specific domain. In an ideal world, each ontology should be linked to a general (or top-level) ontology in order to enable knowledge sharing and reuse [5]. In the domain of the Semantic Web, several ontologies have been developed to describe Web Services.

SPARQL [6] is a query language for RDF. A SPARQL query is represented by a graph pattern to match against the RDF graph. Graph patterns contain triple patterns which are like RDF triples, but with the option of query variables in place of RDF terms in the subject, predicate or object positions. For example, the query composed of the triple pattern *("BRCA1", "is located on", ?chr)* matches the triple described above and returns "chromosome 17" in the variable described chr (variables are identified with the "?" prefix).

In the life sciences community, the use of Semantic Web technologies should be of central importance in a near future. The Semantic Web Health Care and Life Sciences Interest Group (HCLSIG) was launched to explore the application of these technologies in a variety of areas [7]. Currently, several projects have been undertaken. Some works concern the encoding of information using SWL. Examples of data encoded with SWL are MGED Ontology [8], which provides terms for annotating microarray experiments, BioPAX [9], which is an exchange format for biological pathway data, Gene Ontology (GO) [10], which describes biological processes, molecular functions and cellular components of gene products and UniProt [11], which is the world's most comprehensive catalog of information on proteins. Several researches focused on information integration and retrieval [12], [13], [14], [15], [16], [17] and [18] while others concerned the elaboration of a workflow environment based on Web Services[19], [20], [21], [22], [23] and [24].

Regarding the problem of data integration, the application of these technologies faces difficulties which are amplified because of some specificities of biological knowledge.

## Biological data are huge in volume

This amount of data is already larger than what can be reasonably handled by existing tools. In a recent study, Guo and colleagues [25] benchmarked several systems on artificial datasets ranging from 8 megabytes (100,000 declared statements) to 540 megabytes (almost 7 millions statements). The best tested system, DLDB-OWL, loads the largest dataset in more that 12 hours and takes between few milliseconds to more than 5 minutes to respond to the queries. These results, while encouraging, appear to be quite insufficient to be applied to real biological datasets. RDF serialization of the UniProt database, for example, represents more than 25 gigabytes of data. And this database is only one, amongst other data sources that are used on a daily basis by researchers in biology.

## The heterogeneity of biological data impedes data interoperability

Various sources of biological data must be combined in order to obtain a full picture and to build new knowledge, for example data stored in an organism's specific database (such as FlyBase) with results of microarray experiments and information available on related species. However, a large majority of current databases does not use a uniform way to name biological entities. As a result, a same resource is frequently identified with different names. Currently it is very difficult to connect each of these data seamlessly unless they are transformed into a common format with IDs connecting each of them. In the example presented above, the fact that the gene BRCA1 is identified by number "*126611*" in the GDB Human Genome Database [26] and by number "*1100*" in the HUGO Gene Nomenclature Committee (HGNC) [27] requires extra work to map the various identifiers. The Life Sciences Identifier (LSID) [28], a naming standard for biological resources designed to be used in Semantic Web applications should facilitate data interoperability. Unfortunately, at this time, LSID is still not widely adopted by biological data providers.

## The isolation of bio-ontologies complicates data integration

In order to use ontologies at their full potential, concepts, relations and axioms must be shared when possible. Domain ontologies must also be anchored to an upper ontology in order to enable the sharing and reuse of knowledge. Unfortunately, each bio-ontology seems to be built as an independent piece of information in which every piece of knowledge is completely defined. This isolation of bio-ontologies does not enable the sharing and reuse of knowledge and complicates data integration [29].

## A large proportion of biological knowledge is context dependant

Biological knowledge is rapidly evolving; it may be uncertain, incomplete or variable. Knowledge modeling should represent this variation. The function of a gene product may vary depending on external conditions, the tissue where the gene is expressed, the experiment on which this assertion is based or the assumption of a researcher. Databases curators, who annotate gene products with GO terms, use evidence codes to indicate how an annotation to a particular term is supported. Other information that characterizes an annotation can also be relevant (type of experiment, reference to the biological object used to make the prediction, article in which the function is described). This information, which can be considered as a context, constitutes an important characteristic of the assertion which needs to be handled by Semantic Web applications.

## The provenance of biological knowledge is important

In the life science, the same kind of information may be stored in several databases. Sometimes, the contents of information are diverging. In addition to the information itself and the way this information has been generated (metadata encoded by the context), it is also essential, for researchers, to know its provenance [30] (for example, which laboratory or organism has diffused it). Handling the provenance of information is very important in e-science [31]. In bioinformatics, this information is available in several compendia; for example in GeneCards [32] or GeneLynx [33].

A simplified view of the Semantic Web is a collection of RDF documents. The RDF recommendation explains the meaning of a document and how to merge a set of documents into one, but does not provide mechanisms for talking about relations between documents.

Adding the notion of provenance to RDF is envisioned in the future. This topic is currently discussed in a working group called named graphs [34].

Because of these specificities, data integration in the life science constitutes a real challenge.

# Materials and methods

We describe below a use case of biological data integration using Semantic Web technologies. The goal is to build a portal of gene-specific data allowing biologists to query and visualize, in a coherent presentation, various information automatically mined from public sources. The features of the portal, called "Thea online" are similar to other gene portals like GeneCards [32], geneLynx [33], Source [35] or SymAtlas [36]. From the user's point of view, the technical solutions retained to implement the Web site should be totally transparent. Technically, we choose a centralized data warehouse approach in which all the data are aggregated in a central repository.

## Data gathering

We collected various sources of information concerning human genes or gene products. This is an arbitrary choice intended to illustrate the variety of data available and the way these data are processed. Available data are either directly available in SWL, represented in tabular format or stored in tables in relational databases.

Information expressed in SWL concerns protein centric data from UniProt [11], protein interactions data from IntAct [37] (data converted from flat file format into RDF by Eric Jain from Swiss Institute of Bioinformatics) and the structure of Gene Ontology from GO [10] These data are described in two different ontologies. UniProt and IntAct data are described in an ontology called core.owl (available from the UniProt site). GO is a special case in the sense that it is not the definition of instances of an existing ontology, but it is an ontology by itself in which GO terms are represented by classes.

Data represented in tabular format concerns known and predicted protein-protein interactions from STRING [38], molecular interaction and reaction networks from KEGG [39], gene functional annotations from GeneRIFs [40], GO annotations from GOA [41], literature information and various mapping files from NCBI [42].

Information from relational databases is extracted by performing SQL queries. This kind of information concerns Ensembl data [43] which are queried on a MySQL server at address *"ensembldb.ensembl.org"*. A summary of collected data is presented in table 1.

| Source of information | Size of RDF file (in kilobytes) |
|---|---|
| Gene Ontology (at *http://archive.geneontology.org/latest-termdb*) go_daily-termdb.owl.gz | 39,527 |
| GOA (at *ftp://ftp.ebi.ac.uk/pub/databases/GO/goa/HUMAN/*) gene_association.goa_human.gz | 25,254 |
| Intact (RDF description generated by Eric Jain from Uniprot) Intact.rdf | 28,776 |
| Uniprot (at *ftp://ftp.uniprot.org/pub/databases/uniprot/current_release/rdf/*) citations.rdf.gz | 351,204 |
| components.rdf.gz | 6 |
| core.owl | 128 |
| enzyme.rdf.gz | 2,753 |
| go.rdf.gz | 11,541 |
| keywords.rdf.gz | 550 |
| taxonomy.rdf.gz | 125,078 |
| tissues.rdf.gz | 392 |
| uniprot.rdf.gz (human entries only) | 897,742 |
| String (at *http://string.embl.de/newstring_download/*) | |
| protein.links.v7.0.txt.gz (human related entries only) | 388,732 |
| KEGG (at *ftp://ftp.genome.jp/pub/kegg/*) | |
| ftp://ftp.genome.jp/pub/kegg/genes/organisms/hsa/hsa_xrefall.list | 1,559 |
| pathways/map_title.tab | 1 |
| NCBI GeneRIF (at *ftp://ftp.ncbi.nlm.nih.gov/gene/GeneRIF/*) | |
| generifs_basic.gz | 40,831 |
| interactions.gz | 31,056 |
| NCBI mapping files (at *ftp://ftp.ncbi.nlm.nih.gov/gene/DATA/*) | |
| gene2pubmed.gz | 45,092 |
| gene2unigene | 4,036 |
| Mim2gene | 277 |
| NCBI (at *http://eutils.ncbi.nlm.nih.gov/entrez/eutils/efetch.fcgi*) | |
| EFetch for Literature Databases | 317,428 |
| Ensembl MySQL queries (at *ensembldb.ensembl.org*) | 50,768 |
| TOTAL | 2,362,731 |

**Table 1**: List of collected data with the size of the corresponding RDF/OWL specification.

## Data conversion

In the future, when all sources will be encoded in SWL, downloaded data will be imported directly in the data warehouse. But in the meantime, all the data that are not encoded in SWL needed to be converted. Tabular data were first converted in RDF with a simple procedure similar to the one used in YeastHub [44]. Each column which had to be converted in RDF was associated with a namespace that was used to construct the URIs identifying the values of the column (see the section "Principle of URIs encoding" below). The relationship between the content of two columns was expressed in RDF by a triple having the content of the first column as subject, the content of the second column as object and a specified property (figure 1). The conversions from tabular to RDF format were performed by dedicated Java or Python programs. The results obtained by SQL queries, which are composed of set of records, were processed the same way as data in tabular format.

---

a) Protein-protein interaction described in tabular format

9606.ENSP00000046967  9606.ENSP00000334051  600

b) Protein-protein interaction described in RDF

```
<Translation rdf:about="http://www.ensembl.org#ENSP00000046967">
    <interacts_with rdf:ID="SI3" rdf:resource="http://www.ensembl.org#ENSP00000334051"/>
</Translation>
<rdf:Description rdf:about="#SI3">
    <has_score>600</has_score>
</rdf:Description>
```

---

**Fig. 1**: Principle of tabular to RDF conversion. a) a line from the STRING tabular file describing an interaction between a human proteins identified by "*ENSP00000046967*" in the Ensembl database and another protein identified by "*ENSP00000334051*" (in this file, downloaded from STRING, the relationship described in one line is directed, that means that the interaction between "*ENSP00000334051*" and "*ENSP00000046967*" is specified in another line). The reliability of the interaction is expressed by a score of 600 on a scale ranging from 0 to 1000. b) RDF encoding of the same information. The two proteins are represented with URIs and their interaction is represented with the property "*interacts_with*". The triple is materialized by a resource identified by "*SI3*". The score qualifying the reliability of the triple is encoded by the property "*has_score*" of the resource "*SI3*".

## Ontology of generated RDF descriptions

The vocabulary used in generated RDF descriptions is defined in a new ontology called Biowl. Classes (i.e.: Gene, Transcript, Translation) and properties (i.e.: interacts_with, has_score, annotated_with) are defined in this ontology using the namespace URI "*http://www.unice.fr/bioinfo/biowl#*" (for details, see supplementary materials at http://bioinfo.unice.fr/publications/sw_article/).

## Principe of URIs encoding

In the RDF specifications generated from tabular files or from SQL queries, resources are identified with URIs. URIs were built by appending the identifier of a resource in a database to the database URL. For example, the peptide ENSP00000046967 from Ensembl database (accessible at the address *http://www.ensembl.org*) is assigned the URI "*http://www.ensembl.org#ENSP00000046967*" while the gene 672 from NCBI Entrez is assigned the URI "*http://www.ncbi.nlm.nih.gov/entrez#672*".

## Unification of resources

Several lists of mapping between identifiers used in different databases are available on the Web. We used the information from Ensembl, KEGG and the NCBI to generate OWL descriptions specifying the relationship that exists between resources. When two or more resources identify the same object, we unified them with the OWL property *"sameAs"*, otherwise, we used a suitable property defined in the Biowl ontology (figure 2).

```
a)   <GeneProduct rdf:about="http://www.genome.jp/kegg/gene#675">
         <owl:sameAs rdf:resource="http://www.ncbi.nlm.nih.gov/EntrezGene#675"/>
         <biowl:encodes rdf:resource="http://www.ncbi.nlm.nih.gov/Protein#119395734"/>
         <biowl:in_pathway rdf:resource="http://www.genome.jp/kegg/pathway#hsa05212"/>
     </GeneProduct>

b)   <Gene rdf:about="http://www.ncbi.nlm.nih.gov/EntrezGene#675">
         <biowl:cited_in rdf:resource="urn:lsid:uniprot.org:pubmed:1072445"/>
     </Gene>
     <Gene rdf:about="http://www.ncbi.nlm.nih.gov/EntrezGene#672">
         <interacts_with rdf:ID="NI8128" rdf:resource="http://www.ncbi.nlm.nih.gov/EntrezGene#675"/>
     </Gene>
     <Gene rdf:about="http://www.ncbi.nlm.nih.gov/EntrezGene#675">
         <biowl:has_phenotype rdf:resource="urn:lsid:uniprot.org:mim:114480"/>
     </Gene>
     <Gene rdf:about="http://www.ncbi.nlm.nih.gov/EntrezGene#675">
         <in_cluster rdf:resource="http://www.ncbi.nlm.nih.gov/UniGene#Hs.34012"/>
     </Gene>

c)   <Gene rdf:about="http://www.ensembl.org/gene#ENSG00000139618">
         <owl:sameAs rdf:resource="http://www.ncbi.nlm.nih.gov/EntrezGene#675"/>
     </Gene>
```

**Fig. 2**: Description of some of the links between the gene BRCA2, identified with the id 675 at NCBI and other resources. a) descriptions derived from KEGG files. b) descriptions derived from NCBI files. c) descriptions built from the information extracted from Ensembl database.

In the cases where the equivalences between resources are not specified in an existing mapping file, the identification of naming variants for a same resource was manually performed. By looking at the various URIs used to identify a same resource, one can highlight, for example, the fact that the biological process of "cell proliferation" is identified by the URI "http://purl.org/obo/owl/GO#GO_0008283" in GO and by the URI *"urn:lsid:uniprot.org:go:0008283"* in UniProt. From this fact, a rule was built, stating that a resource identified with the URI "http://purl.org/obo/owl/GO#GO_${id}" by GO is equivalent to a resource identified with the URI *"urn:lsid:uniprot.org:go:${id}"* by Uniprot (${id} is a variable that must match the same substring). The GO and Uniprot declarations were then processed with a program that uses the previously defined rule to generate a file of OWL statements expressing equivalences between resources (figure 3).

```
<rdf:Description rdf:about="http://purl.org/obo/owl/GO#GO_0008283">
    <owl:sameAs rdf:resource="urn:lsid:uniprot.org:go:0008283"/>
</rdf:Description>
```

**Fig. 3**: Description of the equivalence of two resources using the owl property "*owl:sameAs*". The biological process of "*cell proliferation*" is identified by the URI "*http://purl.org/obo/owl/GO#GO_0008283*" in GO and by the URI "*urn:lsid:uniprot.org:go:0008283*" in UniProt. This description states that the two resources are the same.

## Ontologies merging

As specified before, in addition to Biowl, we used two other existing ontologies defined by UniProt (core.owl) and GO (go_daily-termdb.owl). These ontologies define different subsets of biological knowledge but are nevertheless overlapping. In order to be useful, the three ontologies have to be unified. There are multiple tools to merge or map ontologies [45] and [46] but they are quite difficult to use and require some user editing in order to obtain reliable results (see the evaluation in the frame of bioinformatics made by Lambrix and Edberg [47]). With the help of the ontology merging tool PROMPT [48] and the ontology editor Protégé [49], we created an unified ontology describing the equivalences between the classes and properties defined in the three sources ontologies. For example, the concept of protein is defined by the class "*urn:lsid:uniprot.org:ontology:Protein*" in Uniprot and the class "*http://www.unice.fr/bioinfo/owl/biowl#Translation*" in Biowl. The unification of these classes is declared in a separate ontology defining a new class dedicated to the representation of the unified concept of protein which is assigned the URI "*http://www.unice.fr/bioinfo/owl/unification#Protein*". Each representation of this concept in other ontologies is declared as being a subclass of the unified concept, as described in figure 4.

```
<owl:Class rdf:ID="http://www.unice.fr/bioinfo/owl/unification#Protein"/>
<owl:Class rdf:about="http://www.unice.fr/bioinfo/owl/biowl#Translation">
    <rdfs:subClassOf rdf:resource="http://www.unice.fr/bioinfo/owl/unification#Protein"/>
</owl:Class>
<owl:Class rdf:about="urn:lsid:uniprot.org:ontology:Protein">
    <rdfs:subClassOf rdf:resource="http://www.unice.fr/bioinfo/owl/unification#Protein"/>
</owl:Class>
```

**Fig. 4**. Unification of different definitions of the concept of Protein (see the text for details).

The same principle is applied for properties by specifying that several equivalent properties are sub properties of a unified one. For example, the concept of name, defined by the property "*urn:lsid:uniprot.org:ontology:name*" in UniProt and the property

"*http://www.unice.fr/bioinfo/owl/biowl#denomination*" in Biowl is unified with the property "*http://www.unice.fr/bioinfo/owl/unification#name*", as shown in figure 5.

```
<owl:DatatypeProperty rdf:ID="http://www.unice.fr/bioinfo/owl/unification#name"/>
<owl:DatatypeProperty rdf:about="urn:lsid:uniprot.org:ontology:name">
    <rdfs:subPropertyOf rdf:resource="http://www.unice.fr/bioinfo/owl/unification#name"/>
</owl:DatatypeProperty>
<owl:DatatypeProperty rdf:about="http://www.unice.fr/bioinfo/owl/biowl#denomination">
    <rdfs:subPropertyOf rdf:resource="http://www.unice.fr/bioinfo/owl/unification#name"/>
</owl:DatatypeProperty>
```

**Fig. 5**. Unification of different definitions of the property name (see the text for details).

One has to note that this is not the aim of this paper to describe a method for unifying ontologies. The unification performed here concerns only obvious concepts like the classes *"Protein"* or *"Translation"* or the properties *"cited_in"* or *"encoded_by"* (for details, see supplementary materials at http://bioinfo.unice.fr/publications/sw_article/).

The unification ontology allows multiples specifications, defined with different ontologies to be queried in a unified way by a system capable of performing type inference based on the ontology's classes and properties hierarchy.

## Data repository

Data collected from several sources which are associated with metadata and organized by an ontology represent a domain knowledge. As we chose a centralized data warehouse architecture, we need to store the set of collected and generated RDF/OWL specifications in a Knowledge Base. In order to be able to fully exploit this knowledge, we need to use a Knowledge Bases System (KBS) [50] capable of storing and performing queries on a large set of RDF/OWL specifications (including the storing and querying of reified statements). It must include reasoning capabilities like type inference, transitivity and the handling of at least these two OWL constructs: "*sameAs*" and "*inverseOf*". In addition, it should be capable of storing and querying the provenance of information.

At the beginning of the project, none of the existing KBS fulfilled these needs. The maximum amount of data handled by existing tools, their querying capabilities and the capabilities to handle contextual information were indeed below our needs (see the benchmark of several RDF stores performed in 2006 by Guo and colleagues [25]). For this reason, we developed and used a KBS specifically designed to answer our needs. Our KBS, called AllOnto is still in active development. It has been successfully used to store and query all the data available on our portal (60 millions of triples including reified statements and the provenance information). At this time, it seems that Sesame version 2.0 (http://www.openrdf.org), released on december 20[th] 2007 has all the features allowing it to be equally used.

## Information retrieval with SPARQL

Triples stored in the KBS, information encoded using reification and the provenance of the assertions can be queried with SPARQL queries. An example of a query allowing retrieving every annotation of protein P38398 associated with its reliability and provenance is given in figure 6.

```
PREFIX   up: <urn:lsid:uniprot.org:uniprot:>
PREFIX unif: <http://www.unice.fr/bioinfo/owl/unification#>
PREFIX  rdf: <http://www.w3.org/1999/02/22-rdf-syntax-ns#>
SELECT
  ?annotation ?reliability ?source
```

```
WHERE
 {
   GRAPH ?source
    { ?r rdf:subject up:P38398 .
      ?r rdf:predicate unif:annotated_by .
      ?r rdf:object ?annot .
      ?r unif:reliability ?reliability
    }
 }
```

**Fig. 6**. SPARQL query used to retrieve annotations of protein P38398. This query displays the set of data representing an annotation, a reliability score and the information source for the protein P38398. The KBS is searched for a triple "*r*" having the resource "*urn:lsid:uniprot.org:uniprot:P38398*" as subject, the resource "*http://www.unice.fr/bioinfo/owl/unification#annotated_by*" as predicate and the variable "*annot*" as object. The value of the property "*http://www.unice.fr/bioinfo/owl/unification#reliability*" of the matched triple is stored in the variable "*reliability*". The provenance of the information is obtained by retrieving the named graph which contains these specifications. The KBS performs *"owl:sameAs"* inference to unify the UniProt protein P38398 with the same resource defined in other databases. It also uses the unified ontology to look for data expressed using sub properties of "*annotated_by*" and "*reliability*".

# Results

## Visualization of collected information about human genome

The Web portal can be accessed at *http://bioinfo.unice.fr:8080/thea-online/*. Entering search terms in a simple text box returns a synthetic report of every available information relative to a gene or gene's product.

Search in Thea-online has been designed to be as simple as possible. There is no need to format queries in any special way or to specify the name of the database a query identifier comes from. A variety of names, symbols, aliases or identifiers can be entered in the text area. For example, a search for the gene BRCA1 and its products can be specified using the following strings: the gene name "*BRCA1*", the alias "*RNF53*", the full sentence "*Breast cancer type 1 susceptibility protein*", the NCBI gene ID "*672*", the UniProt accession number "*P38398*", the OMIM entry "*113705*", the EMBL accession number "*AY304547*", the RefSeq identifier "*NM_007299*" or the Affymetrix probe id "*1993_s_at*".

When Thea-online is queried, the query string is first searched in the KBS. If the string unambiguously identifies an object stored in the base, information about this object is displayed on a Web page. If this is not the case, a disambiguation page is displayed (see figure 7).

**Fig. 7**. Disambiguation page displayed when querying for the string "*120534*". The message indicates that string "*120534*" matches a gene identifier from the Human Genome Database (GDB) corresponding to the Ensembl entry "*ENSG00000132142*" but also matches a gene identifier from KEGG and a gene identifier from NCBI which both correspond to the Ensembl entry "*ENSG00000152219*". A user can obtain a report on the gene he is interested in by selecting the proper Ensembl identifier.

Information displayed as a result of a search is divided in seven different sections: *"Gene Description"*, *"General Information"*, *"Interactions"*, *"Probes"*, *"Pathways"*, *"Annotations"* and *"Citations"*. To limit the amount of data, it is possible to select the type of information displayed by using an option's panel. This panel can be used to choose the categories of information to display on the result page, to select the sources of information to use and to specify the context of some kind of information (this concerns Gene Ontology evidences only at this time).

By performing SPARQL queries on the model, as described in figure 6, the application has access to information concerning the seven categories presented above and some metadata about it. In the current version, the metadata always include the provenance of information, the articles in which an interaction is defined for protein interactions and the evidence code supporting the annotation for gene ontology association. The provenance of information is visualized with a small colored icon (see figure 8). Some exceptions concern information about *"gene and gene products"* and *"genomic location"* which comes from Ensembl and the extensive list of alternative identifiers which are mined from multiple mapping files.

**General Informations**

Genes and Gene Products | **Aliases and Descriptions** | other Identifiers | Probes | Genomic location

- breast cancer 1, early onset isoform BRCA1-delta2-10 *e!*
- breast cancer 1, early onset isoform BRCA1-delta11 *e!*
- RING finger protein 53 **U**
- RNF53 **U**
- breast cancer 1, early onset isoform BRCA1-delta9-11 *e!*
- BRCA1 (Fragment). *e!*
- breast cancer 1, early onset isoform BRCA1-delta9-10 *e!*
- breast cancer 1, early onset *e!*
- breast cancer 1, early onset isoform BRCA1-delta14-17 *e!*
- BRCA1 **U** , *e!*
- breast cancer 1, early onset isoform BRCA1-delta14-18 *e!*
- Breast cancer type 1 susceptibility protein **U**

**Fig. 8**. General information about the gene BRCA1 and its products. Selecting tab "*Aliases and Descriptions*" displays various names and descriptions concerning the gene BRCA1 and its products. Every displayed string is followed by a small icon specifying the provenance of the information: an uppercase red "*U*" for UniProt and a lowercase blue "*e*" followed by a red exclamation mark for Ensembl. Several labels are originating from a single database only (like the string "*breast cancer 1, early onset isoform BRCA1-delta11*" used in Ensembl or "*RING finger protein 53*" used in UniProt) while other labels are common to different databases (like BRCA1 used both in UniProt and Ensembl).

Gene product annotations are displayed as in figure 9. By looking in details at the line describing the annotation with the molecular function "*DNA Binding*", one can see that this annotation is associated with no evidence code in UniProt, with the evidence code "*TAS*" (Traceable Author Statement) in GOA and UniProt and with the evidence code "*IEA*" (Inferred from Electronic Annotation) in GOA and Ensembl. The annotation of the gene product without an evidence code is deduced from the association of the protein with the SwissProt keyword "*DNA-binding*" which is defined as being equivalent to the GO term "*DNA binding*".

**Fig. 9**. Annotations concerning the gene BRCA1 and its products. Selecting tab "*Annotations list*" displays the list of annotations concerning the gene BRCA1 and its products. Every displayed string is followed by a small icon specifying the provenance of the information. The first line, for example, represents an annotation of BRCA1 with the GO term "*regulation of apoptosis*" supported by the evidence code "*TAS*". This information is found in GOA, Ensembl and UniProt. The second line represents an annotation with the GO term "*negative regulation of progression through cell cycle*". This information is found in UniProt with no supporting evidence code and in GOA and Ensembl with the evidence code "*IEA*".

The unification of resources is used to avoid the repetition of the same information. In figure 9, the classification of the protein P38398 with the SwissProt keyword "*DNA-binding*" is considered as being the same information as the annotation with the GO term "*DNA binding*" as the two resources are defined as being equivalent. In figure 10, the unification is used in order to not duplicate an interaction which is expressed using a gene identifier in NCBI and a protein identifier in UniProt and IntAct.

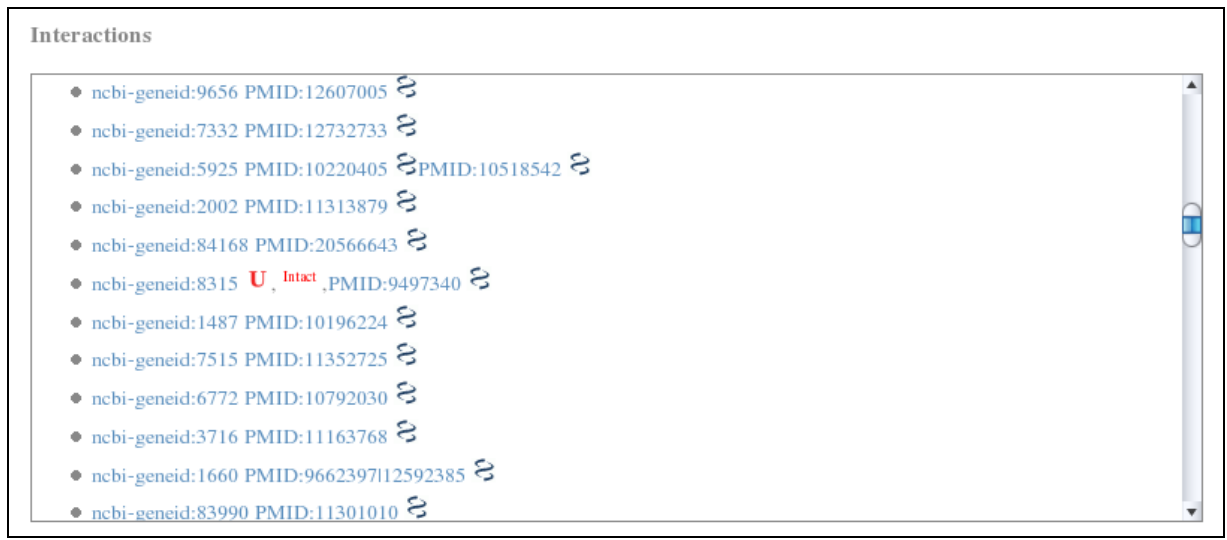

**Fig. 10**. Interactions concerning the gene BRCA1 and its products. Most of the displayed information is coming from NCBI (the NCBI icon is displayed at the end of the lines). When information is available, the Pubmed identifier of the article describing the interaction is given. A line, in the middle of the list displays the same piece of information coming from UniProt, Intact and NCBI. This line is the result of a unification performed on data describing the information in different ways. In UniProt and Intact databases, the protein P38398 is declared as interacting with the protein Q7Z569. In the NCBI interactions file, an interaction is specified with the product of the gene identified by geneID *"672"* and the product of the gene identified by GeneID *"8315"*. The KBS uses the fact that proteins P38398 and Q7Z569 are respectively products of the genes identified at NCBI by the IDs *"672"* and *"8315"* to display the information in a unified way.

## Discussion

Thea-online constitutes a use case of Semantic Web technologies applied to life science. It relies on the use of already available Semantic Web standards (URIs, RDF, OWL, SPARQL) to integrate, query and display information originating from several different sources.

To develop Thea-online, we needed to perform several operations like the conversion of data in RDF format, the elaboration of a new ontology and the identification of each resource with a unique URI. These operations will not be required in the future, when the data will be encoded in RDF. The process of resources mapping will be still needed until the resources are assigned with a unique identifier or that mappings expressed in SWL are available. The same applies to the task of ontology merging that will remain unless ontologies are linked to an upper ontology or until some descriptions of equivalences between ontologies are available.

From the user point of view, the use of Semantic Web technologies to build the portal is not visible. Similar results should have been obtained with classical solutions using for example a relational database. The main impact in the use of Semantic Web technologies concerns the ease of development and maintenance of such a tool. In the present version, our portal is limited to human genes but it can be easily extended to other species. The addition of new kind of data if also facilitated because the KB doesn't rely on a static modelization of the data like in a relational database. To access a new kind of data, one simply has to write or modify a SPARQL request.

Provided that information is properly encoded with SWL, generic tools can be used to infer some knowledge which, until now, must be generated by programming. For example, when searching for a list of protein involved in the response to a thermal stimulus, an intelligent agent, using the structure of Gene Ontology, should return the list of proteins annotated with the term "response to temperature stimulus" but also the proteins annotated "response to cold" or "response to heat". By using the inference capabilities available in AllOnto, the retrieval of the information displayed in each section of a result page is performed with a unique SPARQL query.

Of course, the correctness of the inferred knowledge is very dependent on the quality of the information encoded in SWL. In the current Web, erroneous information can be easily discarded by the user. In the context of the Semantic Web, this filtering will be more difficult because it must be performed by a software agent. Let us take for example the mapping of SwissProt keywords to GO terms expressed in the plain text file spkw2go (http://www.geneontology.org/external2go/spkw2go) . The information expressed in this file is directed: to one SwissProt keyword corresponds one or several GO terms but the reverse is not true. In the RDF encoding of Uniprot, this information is represented in RDF/OWL format with the symmetric property *"owl:sameAs"* (see the file keywords.rdf available from Uniprot RDF site). Thus, the information encoded in RDF/OWL is incorrect but it will be extremely difficult for a program to discover it.

# Conclusion

From this experiment, two main conclusions can be drawn: one which covers the technological issues, the other one which concerns more sociological aspects.

Thea-online is built on a data warehouse architecture [51] which means that data coming from distant sources are stored locally. It is an acceptable solution when the data are not too large and one can tolerate that information is not completely up-to-date with the version stored in source databases. However, the verbosity of SWL results in impressive quantities of data which are difficult to handle in a KBS. An import of the whole RDF serialization of UniProt (25 gigabytes of data) has been successfully performed but improvements are still required in order to deal with huge data-sets. From the technological point of view, the obstacles that must be overcome to fully benefit from the potential of Semantic Web are still important.

However, as pointed out by Good and Wilkinson [52], the primary hindrances to the creation of the Semantic Web for life science may be social rather than technological. There may be some reticences from bioinformaticians to drop the creative aspects in the elaboration of a database or a user interface to conform to the standards [53]. That also constitutes a fundamental change in the way biological information is managed. This represents a move from a centralized architecture in which every actor controls its own information to an open world of inter-connected data which can be enriched by third-parties. In addition, because of the complexity of the technology, placing data on the Semantic Web asks much more work than simply making it available on the traditional Web. Under these conditions, it is not astonishing to note that currently, the large majority of biomedical data and knowledge is not encoded with SWL. Even when efforts were carried out to make the data available on the Semantic Web, most data sources are not compliant with the standards [52], [29] and [14] .

Even though our application works, a significant amount of pre-processing was necessary to simulate the fact that the data were directly available on a suitable format. Other applications on the Semantic Web in life science performed in a similar fashion, using wrappers, converters or extraction programs [44], [54], [55], [56] and [57] . It is foreseeable that, in the future, more data will be available on the Semantic Web, easing the development of increasingly complex and useful new applications. This movement will be faster if information providers are aware of the interest to make their data compatible with Semantic Web standards. Applications, like the one presented in this paper or in others, illustrating the potential of this technology, should gradually incite actors in the life science community to follow this direction.

# Acknowledgements

The author is very grateful to Dr. Richard Christen for critically reading the manuscript.